\documentclass[aps,pre,preprint]{revtex4}
\usepackage{epsfig}
\usepackage{amssymb}
\usepackage{mathrsfs}
\usepackage{latexsym}
\usepackage{wasysym}
\usepackage{pifont}
\usepackage{graphicx}

\begin{document}

\title{Characteristics of Deterministic and Stochastic Sandpile Models
 in a Rotational Sandpile Model}

\author{S. B. Santra, S. Ranjita Chanu and D. Deb}

\affiliation{Department of Physics,\\ Indian Institute of Technology
Guwahati,\\ Guwahati-781039, Assam, India.}
 
\date{\today} 

\begin{abstract}
Rotational constraint representing a local external bias generally has
non-trivial effect on the critical behavior of lattice statistical
models in equilibrium critical phenomena. In order to study the effect
of rotational bias in a out of equilibrium situation like
self-organized criticality, a new two state ``quasi-deterministic''
rotational sandpile model is developed here imposing rotational
constraint on the flow of sand grains. An extended set of new critical
exponents are found to characterize the avalanche properties at the
non-equilibrium steady state of the model. The probability
distribution functions are found to obey usual finite size scaling
supported by negative time autocorrelation between the toppling
waves. The model exhibits characteristics of both deterministic and
stochastic sandpile models.
\end{abstract}

\maketitle

\section{Introduction}

The phenomenon that a class of externally slow driven systems evolves
naturally into a state of no single characteristic size or time
without fine tuning of any parameter is known as self-organized
criticality (SOC)\cite{soc,soc1,nml}. In SOC, the system evolves into
a non-equilibrium steady state characterized by long range
spatio-temporal correlations and power law scaling behavior as
observed in equilibrium critical phenomena\cite{stanley}. The
phenomenon is observed in many natural physical and chemical
processes\cite{exmpls}. Sandpile, a prototypical model, was introduced
by Bak, Tang and Wiesenfeld (BTW)\cite{btw} for studying SOC. The
system here is driven by adding sand grains, one at a time, randomly
to the sites of a regular lattice. In BTW, the model evolves following
certain deterministic rules for distributing grains of a sand column
among all the nearest neighbors equally if the height of a column
attains a predefined critical height. This intermittent burst of
toppling activity is called avalanche. Soon after the introduction of
BTW sandpile model, a stochastic version of the model namely Manna's
stochastic model (MSM)\cite{manna} was proposed. In MSM, sand grains
flow in two randomly selected directions out of four possible
directions on a square lattice after toppling. It was initially
believed that BTW and MSM belong to the same universality
class\cite{piet,guilera,chessa}. However, calculating an extended set
of exponents, introduced by Christensen and Olami\cite{olami}, Ben-Hur
and Biham claimed for the first time that MSM belongs to a new
universality class\cite{benhur}. Different universality class of MSM
has been confirmed later by performing multifractal\cite{stella} and
moment\cite{lubeck} analysis of the avalanche distribution
functions. It is found that the distribution functions of avalanche
properties obey finite size scaling (FSS) in MSM\cite{vespi,chessa}
whereas in BTW, some of them obey multi-scaling\cite{stella,lubeck}.

External bias fields were found to have non-trivial effect on the
critical properties of various lattice statistical models in
equilibrium critical phenomena. For example, in directed self-avoiding
walks (DSAW)\cite{dsaw}, spiral self-avoiding walks (SSAW)\cite{ssaw},
spiral lattice animal (SLA)\cite{sla}, directed percolation
(DP)\cite{dp}, spiral percolation (SP)\cite{sp}, directed spiral
percolation (DSP)\cite{dspr}, etc, the external bias has changed the
critical behavior of these models. The effect of external bias on SOC
is studied so far applying external global directional bias
only. There are two such models, directed sandpile model (DSM)
introduced by Dhar and Ramaswamy\cite{dsp} and directed fixed energy
sandpile (DFES) model introduced by Karmakar and Manna\cite{kar}. In
these models, sand grains flow in a globally preferred direction after
toppling. Application of global bias on the sandpile model introduces
anisotropy in the systems and leads to new universality classes for
DSM\cite{dsp} and DFES\cite{kar}. In contrary, the lattice statistical
models like SLA and SP not only remain isotropic under a local
rotational bias but also exhibit nontrivial critical
behavior\cite{sla,sp}. It is then intriguing to study the effect of a
rotational constraint on the critical properties of sandpile models at
out of equilibrium situation.

In this paper, a new two state ``quasi-deterministic'' rotational
sandpile model (RSM) is introduced imposing rotational constraint on
the flow of sand grains. In this model, a site topples if it exceeds a
predefined critical height and two sand grains flow, one in the
forward direction, the direction from which the last sand grain was
received and the other in a specific rotational direction, say
clockwise with respect to the forward direction. Since the direction
of last sand grain received varies site to site, the rotational
constraint is local in nature. During avalanche, the direction of the
last grain received may change if the toppling sequence is
changed. This introduces certain ``randomness'' or `` internal
stochasticity'' in the model. RSM includes features like mass
conservation, open boundary, local deterministic rule for grain
distribution along with toppling imbalance, certain stochasticity. The
model is studied here numerically in two dimensions ($2D$) and the
critical avalanche properties at the non-equilibrium steady state are
characterized. An extended set of critical exponents are calculated,
nature of the scaling functions are determined, and time auto
correlations between the toppling waves are analyzed. Interestingly,
it is found that some of the critical exponents are similar to that of
BTW but different from that of MSM whereas the scaling functions
follow usual FSS as in MSM. Consequently RSM belongs to a new
universality class. A physical realization of RSM could be a sandpile
on a disk rotating slowly about an axis perpendicular to the plane of
the disk and passing through the center of the disk.

\section{Rotational Sandpile Model}

\begin{figure}
\centerline{\psfig{file=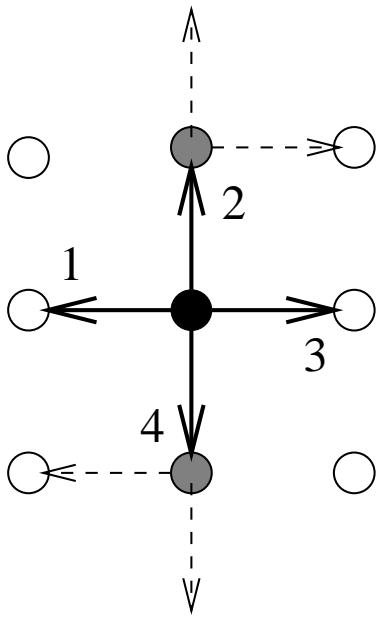,width=0.3\textwidth}}
\medskip 
\caption{\label{model} RSM is demonstrated on a $3\times 3$ square
  lattice. The central black site becomes upper critical first. Arrows
  with numbers $1$-$4$ indicates four possible directions on the
  square lattice. Two sand grains flow along directions $2$ and
  $4$. Consequently, the grey sites become critical. The possible
  directions of flow of sands from the grey sites are indicated by
  dotted arrows. }
\end{figure}

RSM is defined here on a square lattice of size $L\times L$ in two
dimensions ($2D$). A positive integer $h_i$, the height of the sand
column, is assigned to each lattice site. Initially, all $h_i$s are
set to zero. Sand grains are added, one at a time, to randomly chosen
lattice sites and the variable $h_i$ is incremented to $h_i + 1$ if a
sand grain is added to the $i$th site. A site is called active or
critical when the height of a site becomes greater than or equal to a
predefined threshold value $h_c=2$, as in MSM. The active site will
burst into a toppling activity. Toppling of the first active site
initiates an avalanche. This site can be called the origin of the
avalanche. The flow of sand grains is now demonstrated with the help
of Fig.\ref{model}. On the very first toppling of the active site, the
central black circle, two sand grains are given away to two randomly
selected nearest neighbors out of four nearest neighbors on a square
lattice. Note that this is the only externally imposed stochastic step
in the model. The grey circles (sites $2$ and $4$) represent the
recipient sites. As soon as a site receives a sand grain, the
direction $d_i$ from which the grain was received is assigned to it
along with the increment of its height $h_i$. The value of $d_i$ can
change from $1$ to $4$ as there are four possible directions on a
square lattice. The solid arrows in Fig.\ref{model} indicates
different possible directions. As the avalanche propagates, the
direction $d_i$ and height $h_i$ are updated on receiving a sand grain
and the information from which direction the last sand grain was
received is only kept. Now the next active sites with $h_i\ge 2$ in
the avalanche will topple following a deterministic rule. Two sand
grains from a critical site will flow, one in the forward direction,
the direction from which the last grain was received, and the other in
a clockwise rotational direction with respect to the forward
direction. Note that, distribution of sand grains from an active site
depends on the receiving direction of the last sand grain before
toppling which varies site to site. Thus, the rotational constraint
represents a spatially local bias here in RSM in contrast to the
directed bias where the directions of sand flow are globally fixed. If
the grey sites ($2$ and $4$) in Fig.\ref{model} are the new active
sites after receiving sand grains from the central site, the flow of
sands after toppling of the grey sites will be in the forward and
along a rotational direction as indicated by dotted arrows. The
toppling rules for the $i$th active site can be stated as
\begin{equation}
\label{trule}
\begin{array}{cc}
  h_i \rightarrow h_i-2 & \\ h_j \rightarrow h_j+1, & j = d_i
  \hspace{0.2cm} \& \hspace{0.2cm} d_i+1 
\end{array}
\end{equation}
where $d_i$ is the direction from which the last sand grain was
received by the $i$th site. If the index $j$ becomes greater than $4$
it is taken to be $1$. It is important to note that the number of sand
grains outgoing from a site after toppling is not necessarily equal to
the number of sand grains incoming to the same site after toppling of
its nearest neighbors once each. Therefore, there is toppling
imbalance in RSM as in MSM\cite{kms} whereas there is complete
toppling balance in BTW. Note that, toppling imbalance in RSM is due
to the imposition of rotational constraint whereas the same appears in
MSM due to stochasticity in grain distribution. Toppling of an active
site may cause a series of intermittent bursts which constitutes an
avalanche. During an avalanche no sand grain is added. Propagation of
an avalanche stops if all sites of the lattice become under
critical. The avalanche dynamics is studied with open boundary
condition. The number of sand grains remains conserved in the
model. The steady state corresponds to the constant average height of
the sandpile when the current of incoming flux of sand grains is equal
to that of the outgoing flux.

\begin{figure}
\centerline{\psfig{file=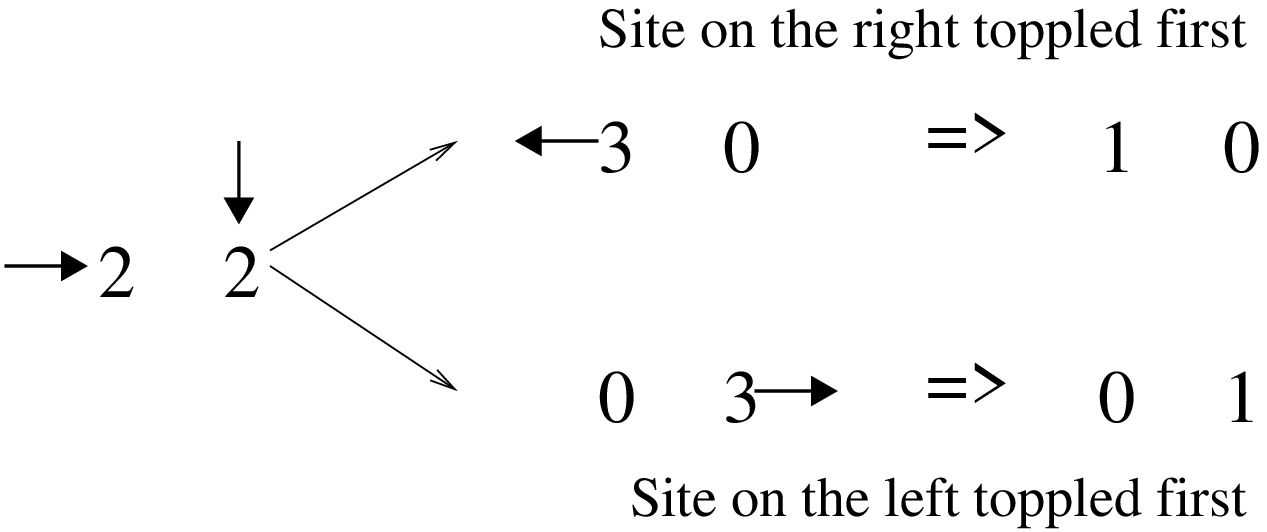, width=0.55\textwidth}}
\medskip
\caption{\label{nonab} Two different final states are obtained
  interchanging the toppling sequence starting from the same initial
  state. The numbers represent the height of the sandpile and the
  associated arrows represent the direction from which the last sand
  grain was received. }
\end{figure}

It should be noted here that the final state in an avalanche depends
on the sequence of toppling of the critical sites due to the
rotational rules considered here. It is demonstrated in
Fig.\ref{nonab} considering two nearest neighbors at the critical
state ($h=2$) at the same time step. The numbers represent the height
of the sand column at that site. The short arrows associated with the
numbers represent the direction from which the last sand grain was
received. It can be seen that reversing the order of toppling, two
different final states are obtained. It has been verified for a larger
lattice that the interchange of toppling sequence leads to different
final configurations starting from the same initial
state. Consequently, the model is non-abelian. Note that toppling of
one site may change the present state (direction from which the last
sand grain was received) of the nearest neighbors. Consequently, the
sand grains will flow in different directions than it was expected
before toppling of its neighbors. Apart from the initial stochastic
step, the imposition of the rotational rule then also introduces some
randomness in the model during time evolution. This can be considered
as ``internal stochasticity'' in the model in contrast to the
externally imposed stochasticity in MSM. This inherent randomness
makes the model not only non-abelian but also
``quasi-deterministic''. Due to the stochastic dynamical rules, MSM
had already been found non-abelian\cite{nonab}. It is also important
to notice that the local correlation in the rotational toppling rule
then can not propagate throughout the avalanche as in BTW because of
the ``internal stochasticity'' in the model.

RSM is thus a new two state, ``quasi-deterministic'', non-abelian
sandpile model under local external rotational bias. Despite few
attempts, the critical behavior of non-abelian sandpile models are
less understood. The non-abelian Zhang model \cite{zhang} belongs to
the same universality class as that of abelian BTW in the limit of
small quantum of energy added to the system\cite{lub}. Directed slope
model is non-abelian\cite{dsm} but it shows the same critical behavior
as that of the abelian directed model\cite{dsp}. A crossover behavior
of critical exponents from a generalized Zhang model to that of BTW
model depending upon a non universal parameter $p$, the probability of
sand flow in a given direction, is shown by Biham {\em et
al}\cite{biham}. On one hand, RSM has features of mass conservation,
open boundary, local deterministicity in gain distribution to the
neighborhood after toppling as that of BTW, on the other hand, it has
features like toppling imbalance, certain stochasticity and
non-abelianity as that of MSM. It is then interesting to characterize
the critical properties of RSM which has microscopic features of both
BTW and MSM. Below, the avalanche properties at the steady state are
characterized in three different ways: $(i)$ calculating an extended
set of critical exponents, $(ii)$ performing moment analysis of the
probability distribution functions and $(iii)$ coarsening an avalanche
into a series of toppling waves.

\section{Results and Discussion}
\subsection{The steady state}
\begin{figure}
\centerline{\psfig{file=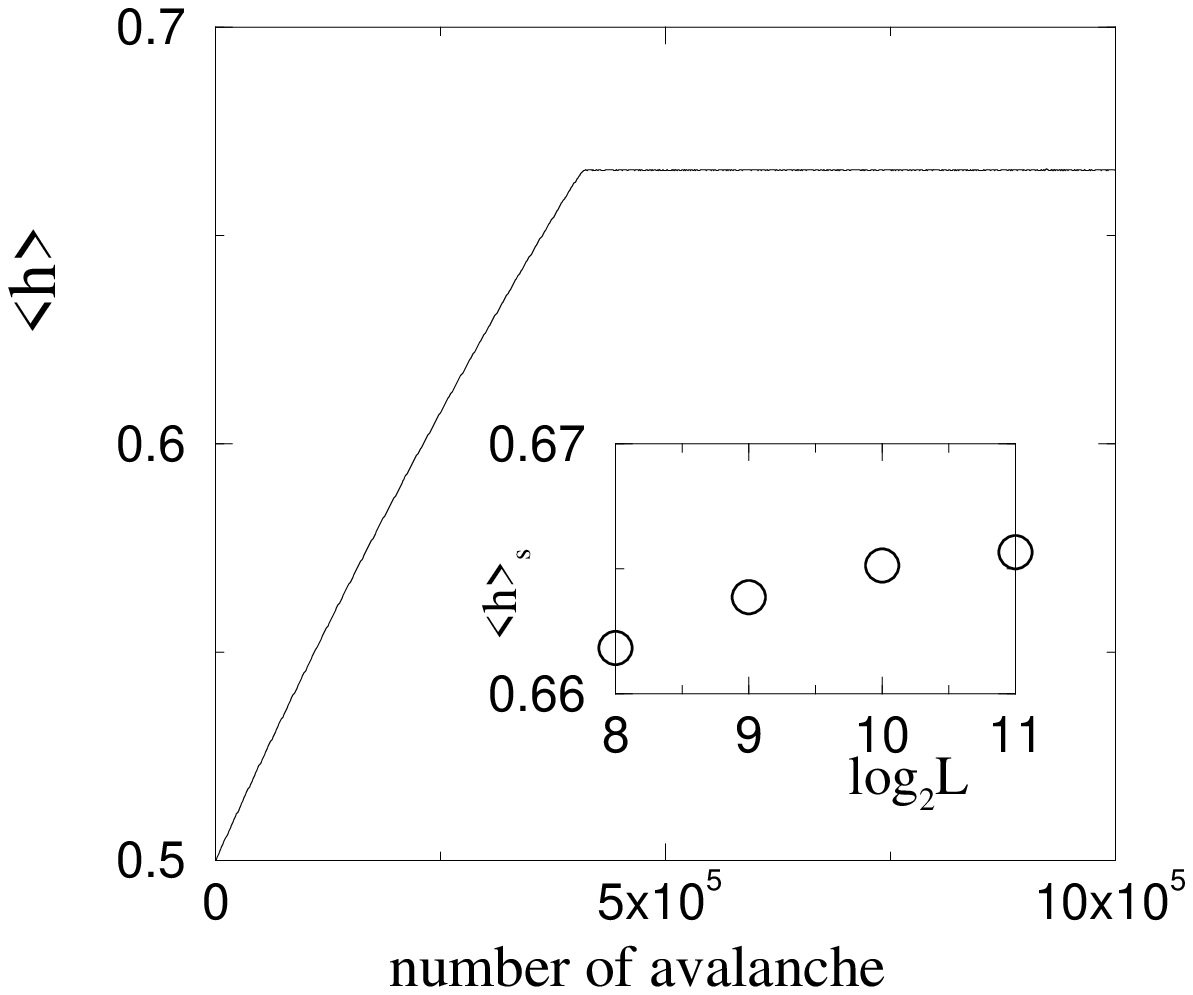, width=0.65\textwidth}}
\medskip
\caption{\label{sat} Plot of average height $\langle h \rangle$
  against the number of avalanches. The value of $\langle h \rangle$
  remains constant over a large number of avalanches and changes
  slightly with the system size $L$. Dependence of the saturated
  average height $\langle h \rangle_s$ on the system size $L$ is shown
  in the inset. }
\end{figure}

The non-equilibrium steady state is defined by the constant average
height of the sandpile at which the current of influx of sand grain to
the system is equal to the current of out flux of the same at the open
boundary. In order to identify the steady state, the average height
\begin{equation}
\label{ah}
\langle h \rangle =\frac{1}{L^2} \sum_{i=1}^{L^2}h_i
\end{equation}
has been measured generating a large number of avalanches. The average
height $\langle h \rangle$ is plotted against the number of avalanches
upto $10^6$ in Fig.\ref{sat} for the system size $L=2048$. It can be
seen that a constant average height $\langle h \rangle$ is achieved
and it remains constant over a large number of avalanches. For smaller
lattice sizes the steady states are reached by smaller number of
avalanches. A slight variation of the average height with the system
size is observed. The values of $\langle h \rangle$ against the system
size $L$ is shown in the inset. In order to characterize the physical
properties of the avalanches occurred at the non-equilibrium steady
state, simulations have been performed on the square lattice of sizes
$L=128$ to $L=2048$ in multiple of $2$. First $10^6$ avalanches were
skipped to achieve the steady state. Extensive data collection have
been made for each lattice size for averaging, ranging from $32\times
10^6$ avalanches for $L=128$ down to $2\times 10^6$ avalanches for
$L=2048$ in ten configurations. In each configuration, the initial
$10^5$ avalanches are neglected again on the steady state before
collecting data. It should be mentioned here that due to the
rotational constraint the lifetime of an avalanche in RSM is much
higher in comparison to that in other models. Generation of large
number of avalanche then requires huge computer time in RSM.

\subsection{Avalanche cluster Morphology}
\begin{figure}
\centerline{\hfill \psfig{file=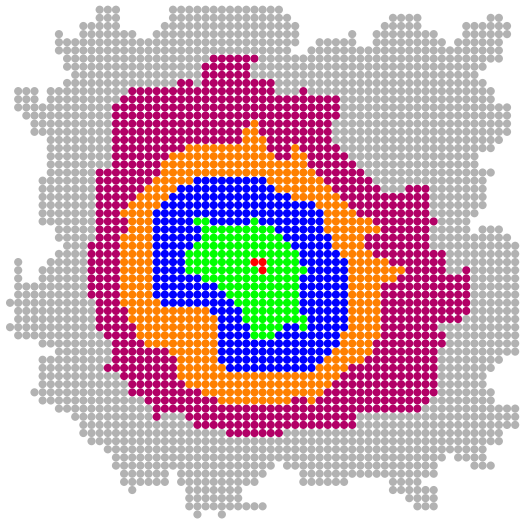,width=0.45\textwidth}
  \hfill\hfill \psfig{file=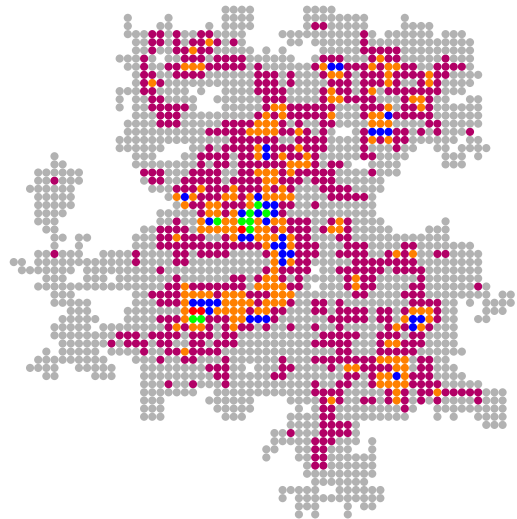,width=0.45\textwidth} \hfill }
\medskip
\centerline{\hfill $(a)$ BTW \hfill\hfill $(b)$ MSM \hfill }
\medskip
  \centerline{\psfig{file=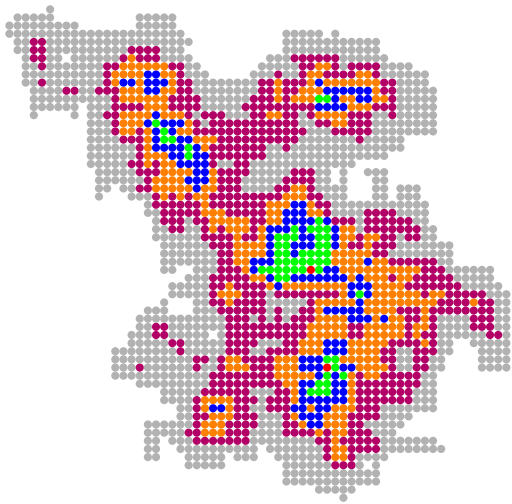,width=0.45\textwidth} }
\medskip
\centerline{\hfill $(c)$ RSM \hfill }
\medskip
\caption{\label{avch} (Color online) Typical avalanches generated at
  the steady state are shown for BTW $(a)$, MSM $(b)$, and RSM $(c)$
  on a square lattice of size $64\times 64$. Avalanches are generated
  dropping sand grains at the central site of the lattice. Maximum
  number of toppling occurs in each clusters is $21$. Different colors
  chosen are: red for $21$, green for $20$-$17$, blue for $16$-$13$,
  orange for $12$-$9$ , magenta for $8$-$5$, and grey for $4$-$1$
  toppling numbers. White space inside the avalanche corresponds to
  the sites that did not topple at all during the avalanche. Avalanche
  cluster of RSM has characteristics of both BTW and MSM.}
\end{figure}

A comparison of the morphology of avalanche clusters in the steady
state of different sandpile models is made here. Typical large
avalanche clusters of BTW, MSM and RSM obtained in their respective
steady states are shown in Fig.\ref{avch} (color online). The
avalanche clusters are generated on a square lattice of size $64\times
64$ dropping sand grains one at a time at the center of the
lattice. The clusters shown here have $21$ maximum number of toppling
in each and it is represented by the red color. Different colors
correspond to different number of toppling of sites in an avalanche
as: red for $21$, green for $20$-$17$, blue for $16$-$13$, orange for
$12$-$9$, magenta for $8$-$5$, and grey for $4$-$1$ toppling
numbers. White spaces inside the avalanche correspond to the sites
that did not topple at all during the avalanche. There are few things
to notice. First, RSM avalanche cluster is different from that of both
BTW as well as MSM. BTW avalanche cluster consists of concentric zones
of lower and lower number of toppling around a single maximal toppling
zone\cite{comp} whereas avalanche cluster of MSM is random in
nature\cite{benhur}. RSM avalanche cluster neither fully consists of
concentric zones as in BTW nor it is totally random. It is important
to note that the local correlation in rotational constraint does not
lead to long range correlation generating BTW like correlated
structure. Second, there are more than one red zones (maximum number
of toppling zones) in the avalanche cluster of RSM in
Fig.\ref{avch}($c$) as in the avalanche cluster of MSM
(Fig.\ref{avch}($b$)) whereas there is only one red zone at the center
of the avalanche cluster of BTW (Fig.\ref{avch}($a$)). It is then
possible to have several maximal toppling zones in an avalanche in RSM
whereas in BTW always one maximal toppling zone
appears\cite{comp}. RSM avalanche looks like superimposition of
several BTW like structures around different maximal toppling
zones. Occurrence of several maximal toppling zones is also a common
feature in MSM avalanche\cite{comp,benhur}. Third, though the sand
grains are added at the central site of the lattice the maximal
toppling zones appear at arbitrarily different places in RSM as well
as in MSM. In BTW, the central zone correspond to the maximal toppling
zone\cite{comp}. Fourth, BTW avalanche clusters are compact without
holes or no toppling regions inside an avalanche (see
Fig.\ref{avch}($a$)). On the other hand, there are several holes
appear in MSM as seen in Fig.\ref{avch}($b$). RSM avalanche cluster is
almost compact with few holes here and there. The appearance of holes
in MSM is due to the stochastic rule of sand distribution and the same
in RSM is due the rotational rule of sand distribution. Note that
holes of a single site also appear in the avalanche clusters of
RSM. Apparently it seems generation of single sited hole is forbidden
by the rotational rule. Single sited holes only could appear at the
termination point of two different branching of toppling of a previous
site. An avalanche can be considered as a branching process since
toppling of a site can make more than one neighbors
critical\cite{ssms}. The RSM avalanche cluster therefore has
properties of both the deterministic BTW and stochastic MSM. It is
then expected that the critical avalanche properties in the steady
state will show a mixed behavior. Note that, the avalanche clusters
are isotropic in space in RSM as in both BTW and MSM. However, the
avalanche clusters are anisotropic in the cases of DSM and DFES where
two correlation lengths are required to characterize their spatial
extensions.

\subsection{Criticality and Power laws}
\begin{figure}
\centerline{\psfig{file=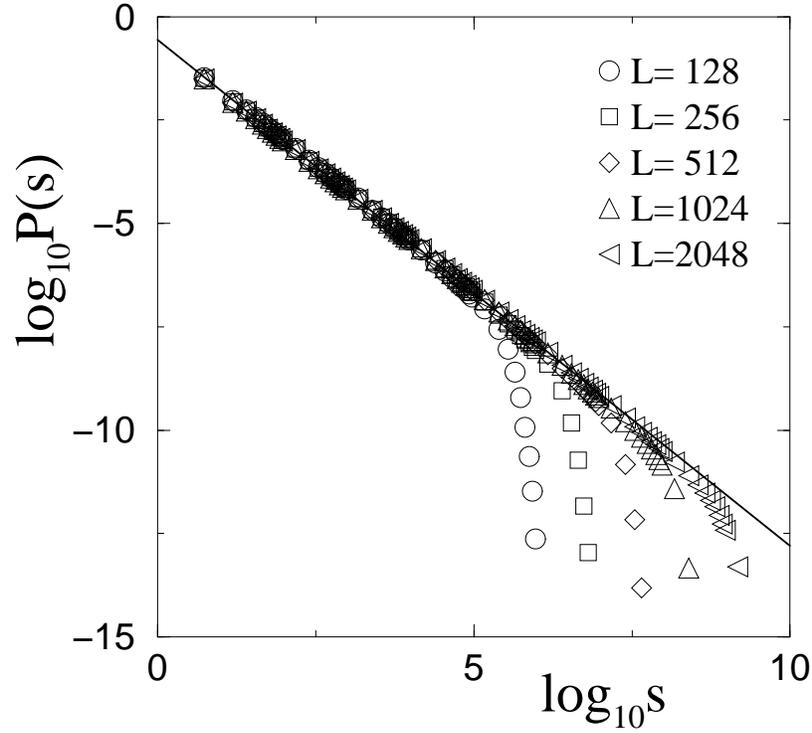, width=0.65\textwidth}}
\medskip
\caption{\label{pds} $P(s)$ is plotted against $s$ for different system
  sizes $L$: $L=128$ ($\Circle$), $L=256$ ($\Box$), $L=512$
  ($\Diamond$), $L=1024$ ($\triangle$), $L=2048$
  ($\triangledown$). The solid lines show the best fitted part and the
  slope correspond to $\tau_s=1.224\pm 0.005$. Toppling size
  distribution has power law correlation at the steady state.}
\end{figure}

\begin{figure}
\centerline{\psfig{file=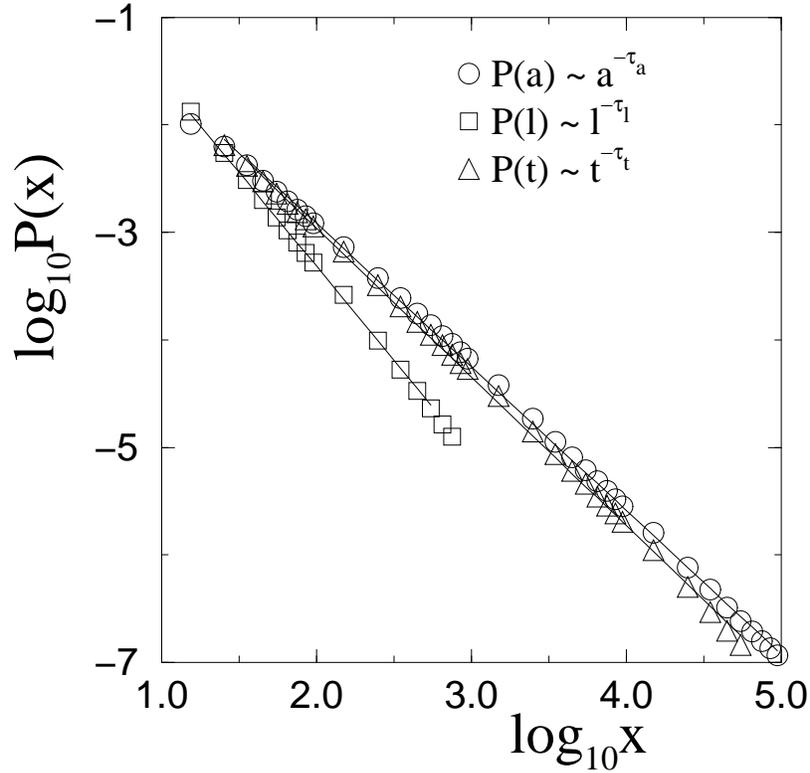, width=0.65\textwidth}}
\medskip
\caption{\label{pdatl} Plot of probability distributions of avalanche
  area $P(a)$ ($\Circle$), lifetime $P(t)$ ($\triangle$) and extension
  of avalanche $P(l)$ ($\Box$) against the corresponding variables
  $a,t$ and $l$ for $L=2048$. Reasonable power law distributions are
  obtained for all three properties. The solid lines show the best
  fitted parts and the slopes correspond to the respective exponents
  $\tau_a=1.334\pm 0.005$, $\tau_t=1.389\pm 0.005$ and
  $\tau_l=1.667\pm 0.007$. Errors are least square fit error taking
  into account of statistical errors of each data points. }
\end{figure}

In order to characterize different physical properties of the
avalanches occurred at the steady state, different quantities like
toppling size $s$ of an avalanche, avalanche area $a$, lifetime $t$ and
spatial extension $l$ are measured. Toppling size $s$ is defined as
the total number of toppling occurred in an avalanche. Avalanche area
$a$ is equal to the number of distinct sites toppled in an
avalanche. Lifetime $t$ of an avalanche is taken as the number of
parallel updates to make all the sites under critical. Spatial
extension $l$ of an avalanche is given by $l^2 = 2\sum_{i=1}^a({\bf
r}_0 - {\bf r}_i)^2/a$ where ${\bf r}_0 = \sum_{i=1}^a{\bf r}_i/a$,
${\bf r}_i$ is the position vector of the distinct sites toppled. The
related critical exponents are estimated determining the probability
distributions of all these properties ($s,a,t$ and $l$). The
probability distribution function of an avalanche related quantity $x$
at the steady state of a given system size $L$ is expected to obey
power law behavior given by
\begin{equation}
\label{pd}
P(x,L) \sim x^{-\tau_x}{\sf f}(x/L^{D_x})
\end{equation}
where $\tau_x$ is the corresponding critical exponent and $x$ stands
for $s,a,t$ and $l$. ${\sf f}(x/L^{D_x})$ is the finite size scaling
function and $D_x$ is called a capacity dimension. In the
$L\rightarrow \infty$ limit, the scaling function ${\sf f}(0)$ become
a constant and the power law behavior given in Eq.\ref{pd} can be
approximated as $P(x) \sim x^{-\tau_x}$. The corresponding exponents
$\tau_x$ can be estimated from the slope of the best fitted straight
line through the data points in log-log scale. Data are collected in
bins of interval of $10$s, $100$s, $1000$s and so on. Finally the data
are normalized by the bin widths. In Fig.\ref{pds}, the probability
distribution $P(s)$ of the toppling size $s$ is plotted for different
system sizes, $L=128$ to $L=2048$ in multiple of $2$. It can be seen
that the toppling size distribution has the same power law behavior
for different system size $L$ with a cutoff that increases with
$L$. Since there is no typical toppling sizes of an avalanche at the
non-equilibrium steady state, RSM then exhibits self-organized
criticality. The solid line represents the best fitted straight line
between the data points of $L=2048$ with a slope of $1.224\pm
0.005$. The error is due to the least square fitting taking into
account of statistical error of each data point. The slopes obtained
from the best fitted part of data for other values of $L$ remain
within this error bar. In order to extract the critical exponents
related to other avalanche properties the same procedure has been
followed. However, in Fig.\ref{pdatl} the probability distribution
$P(x)$ is plotted against $x$ only for $L=2048$ where $x$ corresponds
to area $a$ (circles), lifetime $t$ (triangles) and spatial extension
$l$ (squares). It can be seen that the probability distributions
$P(x)$ follow reasonable power law behavior for each property $x$. The
solid lines represent the best fitted straight line through the data
points. The values of the associated critical exponents $\tau_x$ are
obtained from the slopes of the best fitted straight lines as $\tau_a
= 1.334\pm 0.005$, $\tau_t= 1.389\pm 0.005$ and $\tau_l= 1.667\pm
0.007$. The error bars quoted here are the least square fit errors
taking into account of statistical error of each data point. A
comparison of the values of the exponents obtained here is made with
that of BTW and MSM in Table \ref{table1}. The values of the exponents
for BTW and MSM are taken from
Ref.\cite{benhur,usadel}. Interestingly, the toppling size exponent
$\tau_s$ and the lifetime exponent $\tau_t$ are different whereas
$\tau_a$ and $\tau_l$ are almost the same as that of BTW. The
disagreement of the lifetime and toppling size distribution exponents
with the corresponding BTW exponents can be accounted by the fact that
in RSM the avalanche waves generally have a spiraling nature around
several maximal toppling zones within the avalanche cluster and as a
consequence it will take longer time and large number of toppling for
an avalanche to die away than that in BTW where a single maximal
toppling zone occurs and the toppling wave propagates outwardly. On
the other hand, in comparison to MSM, most of the exponents are found
different. Thus, from the point of view of power law correlations,
some of the avalanche properties are similar to that of BTW but
different from MSM. Note that $\tau_s=2-1/\tau_a$, conjectured by
Majumder and Dhar\cite{mjdh}, is satisfied in case of MSM but it is
not valid for BTW. It can be seen that the conjecture is just outside
the error bar here in case of RSM. The expected value of $\tau_s$ in
RSM from the conjecture is $\approx 1.25$ close to the obtained value
$1.224\pm 0.005$.

\begin{figure}
\centerline{\psfig{file=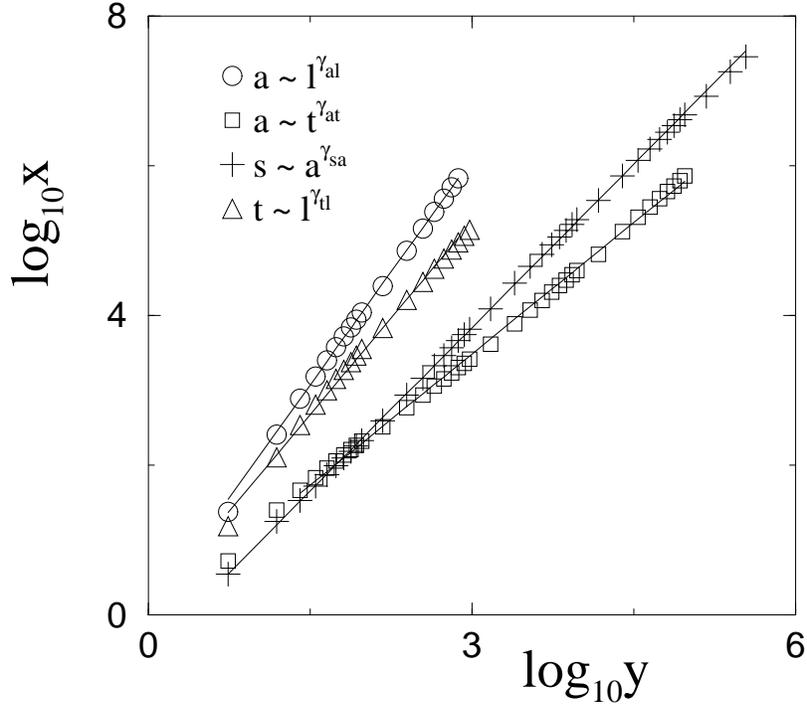, width=0.65\textwidth}}
\medskip
\caption{\label{ealt} Plot of conditional probabilities of avalanche
properties: toppling ($s$) versus area ($a$) ($+$), area ($a$) versus
extension ($l$) ($\circ$), area ($a$) versus time ($t$) ($\Box$), and
time ($t$) versus length ($l$) ($\triangle$). The solid lines show the
best fitted straight line parts. Corresponding exponents are found as:
$\gamma_{sa} = 1.34\pm 0.01$, $\gamma_{at} = 1.167\pm 0.005$,
$\gamma_{al} = 2.002\pm 0.002$ and $\gamma_{tl}=1.713\pm
0.015$. Errors are least square fit error.}
\end{figure}

Since the avalanche properties are related to each other, conditional
expectation values are defined as introduced by Christensen {\em et
al}\cite{chris}. The conditional expectation value of an avalanche
property $x$ when another property is exactly equal to $y$ is defined
as
\begin{equation}
\label{cev}
\langle x(y)\rangle = \sum_x xP(x,y) 
\end{equation}
where $P(x,y)$ is the probability to find a property $x$ when the
other property is exactly equal to $y$ for a given system size $L$. In
the steady state, the expectation values scale with its argument as
\begin{equation}
\label{gamma}
\langle x(y)\rangle \sim y^{\gamma_{xy}}
\end{equation}
where $\gamma_{xy}$ is a critical exponent. Four expectation values
$\langle s(a)\rangle \sim a^{\gamma_{sa}}$, $\langle a(t)\rangle \sim
t^{\gamma_{at}}$, $\langle a(l)\rangle \sim l^{\gamma_{al}}$, and
$\langle t(l)\rangle \sim l^{\gamma_{tl}}$ are calculated on a square
lattice of size $L=2048$ and their scaling behavior are determined. In
Fig.\ref{ealt}, the expectation values are plotted against their
arguments in order to evaluate the exponents $\gamma_{xy}$. From the
slope of the best fitted straight lines, the values of the exponents
are estimated as: $\gamma_{sa}= 1.453\pm 0.003$, $\gamma_{at}=
1.167\pm 0.005$, $\gamma_{al}=2.002\pm 0.002$ and $\gamma_{tl}=
1.715\pm 0.005$. The values of $\gamma_{xy}$ in DSM\cite{benhur} are
smaller in comparison to the values obtained here. In Table
\ref{table1}, the values of $\gamma_{xy}$ are compared with that of
BTW and MSM.  There are few things to notice. First, $\gamma_{sa}$ is
found greater than one and a relevant exponent. This is expected
because in this model, a site topples many times in an avalanche due
to rotational constraint. Second, the exponent $\gamma_{al}$ is found
$\approx 2$ since the avalanche clusters are almost compact with a few
holes here and there. Third, the value of the dynamical exponent
$\gamma_{tl}$ is the highest in RSM and it is lowest in BTW. Because,
due to the rotational constraint the sand grains rotates around
several maximal zones inside the avalanche and take longer time to
complete an avalanche. Fourth, according to the scaling function form
given in Eq.\ref{gamma}, the exponents should satisfy the scaling
relation $\gamma_{xz} = \gamma_{xy}\gamma_{yz}$. It can be seen that
the scaling relation $\gamma_{al}=\gamma_{at}\gamma_{tl}$ is satisfied
within error bars. Fifth, the values of $\gamma_{sa}$, $\gamma_{at}$
and $\gamma_{tl}$ are found different from that of BTW as well as MSM
except $\gamma_{al}$. Finally, a set of scaling relations between the
probability distribution exponents $\tau_x$ and the exponents
$\gamma_{xy}$ describing the conditional expectation values of the
avalanche properties can be obtained from the following identity
\begin{equation}
\label{idnty}
\int \langle x(y)\rangle P(y)dy = \int \langle x(z)\rangle P(z)dz
\end{equation}
which would be satisfied by any set of three stochastic variables $x$,
$y$ and $z$. Using this identity and the relation $\gamma_{xz} =
\gamma_{xy}\gamma_{yz}$, the following scaling relation can be obtained
\begin{equation}
\label{sclr}
\gamma_{xy} = (\tau_y-1)/(\tau_x-1)
\end{equation}
The above scaling relation is satisfied within error bars for $x,y\in
\{s,a,l,t\}$ . Thus, the extended set of exponents obtained here in
RSM from both power law analysis and conditional probabilities are
consistent with the scaling relations. Note that the values of the
critical exponents obtained here should remain invariant under the
reversal of rotational symmetry. Though some of the exponents are
close to that of BTW, RSM belongs to a new universality class because
the extended set of exponents are not identical. Note that, DSP and
DEFS already belongs to different universality classes because of
their anisotropic character. Directed models show a continuous phase
transition from an absorbed phase to an active
phase\cite{dsp,dsp1,kar}.

\subsection{Moment Analysis and Finite Size Scaling}
\begin{figure}
\centerline{\psfig{file=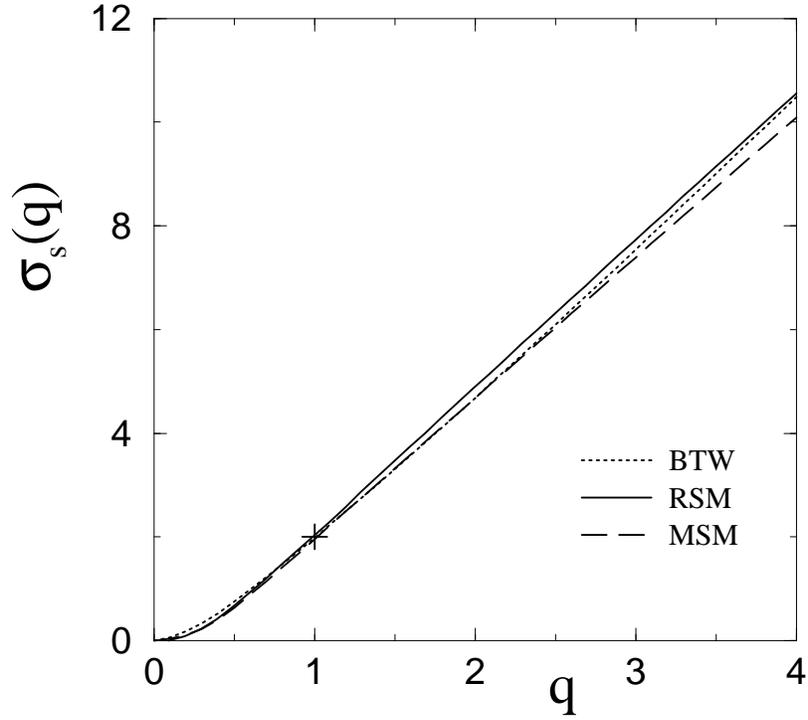, width=0.65\textwidth}}
\medskip
\caption{\label{sq1} Plot of $\sigma_s(q)$ versus $q$. Data of RSM
(solid line) are compared with that of BTW (dotted line) and MSM
(dashed line). $\sigma_s(q)$ of RSM is not identical with that of
either BTW or MSM. Plus sign corresponds to the coordinate $q=1$ and
$\sigma_s(1)=2$. }
\end{figure}

\begin{figure}
\centerline{\psfig{file=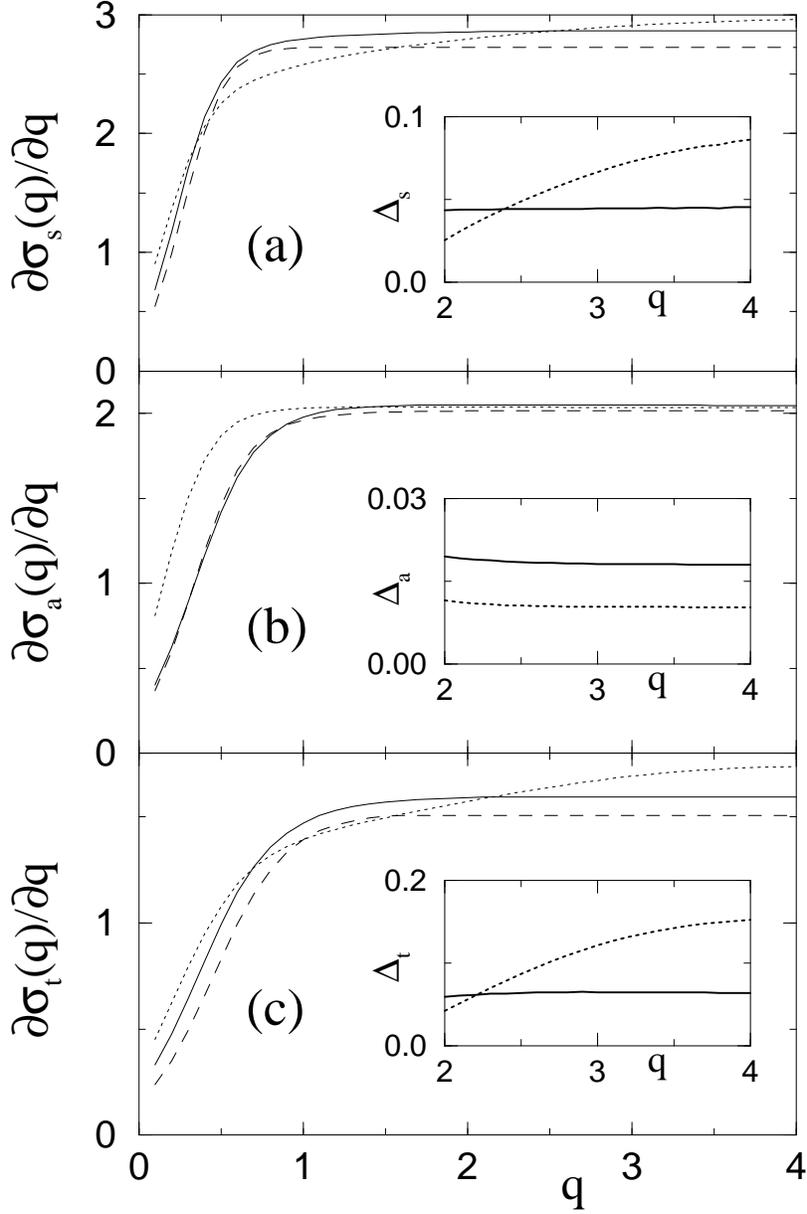, width=0.65\textwidth}}
\medskip
\caption{\label{mdsat} Plot of $\partial\sigma_x(q)/\partial q$ versus
$q$ where $x\in \{s,a,t\}$ in $(a)$, $(b)$ and $(c)$
respectively. Data of RSM (solid line) are compared with that of BTW
(dotted line) and MSM (dashed line). The rate of change of $\sigma_x$s
remain constant for RSM as in the case of MSM. In case of BTW,
$\partial\sigma_s(q)/\partial q$ and $\partial\sigma_t(q)/\partial q$
do not saturate with the moment $q$. The relative increment of the
rates $\Delta_x$ with respect to MSM are plotted against the moment
$q$ in the inset of respective plots. Dotted lines represent the
relative change of rates of BTW with respect to MSM and the solid
lines represent the same for RSM with respect to MSM. }
\end{figure}

Now the avalanche properties are analyzed to understand the nature of
the scaling functions, FSS or multi-scaling, following the method of
moment analysis\cite{stella,lubeck,kms}. The probability distribution
$P(x,L)$ of an avalanche property $x$ in a finite system of size $L$
is expected to obey a scaling function form as given in Eq.\ref{pd},
$P(x,L) \sim x^{-\tau_x}{\sf f}(x/L^{D_x})$ where $D_x$ is called a
capacity dimension as already mentioned. The finite system size $L$
causes a cutoff of the probability distributions at $x_{max} \sim
L^{D_x}$. The $q$-moments of a property $x$ is defined as
\begin{equation}
\label{qmnts}
\langle x^q \rangle = \int_0^{x_{max}} x^{q}P(x,L)dx \sim
L^{\sigma_x(q)}
\end{equation}
where $\sigma_x(q)=(q+1-\tau_x)D_x$. Thus, if the probability
distributions obey FSS then the moment exponent $\sigma_x(q)$ should
have a constant gap between two successive values,
$\sigma_x(q+1)-\sigma_x(q)=D_x$. On the other hand, if they obey
multi-scaling, $\sigma_x(q)$ should have a continuous dependence on
$q$. For a given $q$, the value of $\sigma_x$ has been obtained from
the slope of the plot of $\log\langle x^q(L)\rangle$ versus $\log L$
changing lattice size from $L=128$ to $2048$ in multiple of $2$.  The
least square fit error for each $\sigma_x$ value is found $\pm
0.01$. A sequence of exponents $\sigma_x(q)$, $x\in \{s,a,t\}$, is
obtained for $400$ values of $q$ between $0$ and $4$. In
Fig.\ref{sq1}, the exponent related to the average toppling size
$\sigma_s(q)$ is plotted against the moment $q$. An important point to
note here is that the value of $\sigma_s(1)$ is $\approx 2$ in all
three models. The average toppling size varies with the system size as
$\langle s \rangle \sim L^{\sigma_s(1)}$ where $s$ is equivalent to
the number of steps by a random walker required to reach the lattice
boundary\cite{ssms}. Thus, the value of $\sigma_s(1)\approx 2$
represents the diffusive character of the model, one of the
characterizing features of the sandpile models. The diffusive behavior
of RSM is also consistent with the fact that spiral random walks are
diffusive\cite{bill}. In order to compare the results of RSM with that
of BTW and MSM, a sequence of exponents $\sigma_x(q)$ are also
obtained for these two models and plotted in the same figure. It can
be seen that the variation of $\sigma_s(q)$ with $q$ in Fig.\ref{sq1}
is not identical with either BTW or MSM. Further analysis of this
sequence of exponents is then needed in order to understand the nature
of the scaling function. The slopes $\partial\sigma_x/\partial q$ are
then estimated using finite difference method. If the probability
distributions obey multi-scaling, the rate of change of $\sigma_x(q)$
with $q$ should not be a constant. In Fig.\ref{mdsat},
$\partial\sigma_x/\partial q$s are plotted against the moment $q$ for
$x\in \{s,a,t\}$ and compared with that of BTW and MSM.  In
Fig.\ref{mdsat}, the solid line represents the data of RSM, the dashed
line represents the data of MSM and the dotted line represents the
data of BTW. The rate of change of $\sigma_x(q)$ with respect to $q$
for all three properties $x\in \{s,a,t\}$ in RSM are different from
that of both BTW and MSM and remain unchanged in higher moments as in
the case of MSM. In BTW, the derivatives corresponding to toppling
size $s$ and lifetime $t$ do not saturate with the moment $q$. A
comparison of variation of $\partial\sigma_x/\partial q$ with $q$ in
BTW and RSM with respect to MSM can be made. In order to have a
comparative study a quantity $\Delta_{x,m}(q)$ is defined as
\begin{equation}
\label{dqm}
\Delta_{x,m}(q)
=\left|1-\frac{\sigma'_{x,m}(q)}{\sigma'_{x,MSM}(q)}\right|
\end{equation}
where $\sigma'(q)=\partial\sigma/\partial q$ and $m$ stands for the
models BTW or RSM. $\Delta_{x,m}(q)$ is plotted against $q$ in the
inset of corresponding plots. The dotted line corresponds to
$\Delta_{x,BTW}$ and the solid line corresponds to
$\Delta_{x,RSM}$. It can be seen that the value of $\Delta_{x,RSM}(q)$
remains constant with respect to MSM whereas in case of BTW, data
corresponding to toppling size and lifetime increases slowly with
respect to MSM. Thus, the scaling functions of the avalanche
properties in RSM follow FSS as in MSM rather than multi-scaling as in
BTW.

The values of the capacity dimensions $D_x$ can be calculated taking
the large $q$ limit of $\partial\sigma_x(q)/\partial q$. The values
obtained are: $D_s=2.86\pm 0.01$, $D_a=2.03\pm 0.01$ and $D_t=1.60\pm
0.01$. The errors are due to the finite difference method adopted for
differentiation of the $\sigma(q)$ sequence. The value of $D_l$ is
trivially equal to $1$ because $l_{max}\sim L$. Since $\int P(x)dx =
\int P(y)dy$ for a given system size $L$, it can be shown that
$D_x/D_y = \gamma_{xy}$. Taking $D_l=1$, one should have $D_x =
\gamma_{xl}$. For RSM, the values of $D_x$ and $\gamma_{xl}$ are found
close. Since $\partial\sigma_s(q)/\partial q$ and
$\partial\sigma_a(q)/\partial q$ do not saturate in BTW, the
corresponding capacity dimensions are not possible to
estimate. However, the capacity dimension $D_a$ in all three models
are found $\approx 2$ as it is expected. The values of $D_s$ and $D_t$
in MSM (as in Ref.\cite{chessa}) and RSM differ slightly. Note that
$D_s(2-\tau_s)\approx 2.22$ is slightly higher than
$\sigma_s(1)\approx 2$ because $q=1$ remains in the nonlinear regime
in RSM.

\begin{figure}
\centerline{\psfig{file=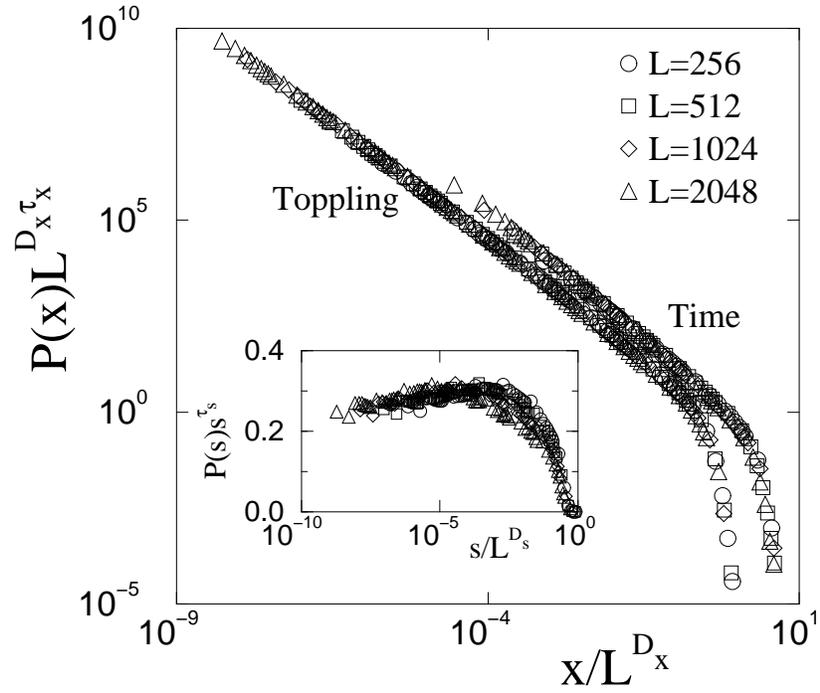, width=0.65\textwidth}}
\medskip
\caption{\label{fss} Plot of the scaled distribution
$P(x)L^{D_x\tau_x}$ against the scaled variable $x/L^{D_x}$ for
$x\in\{s,t\}$. System sizes are taken as $L=256$ ($\Circle$), $L=512$
($\Box$), $1024$ ($\Diamond$) and $2048$ ($\triangle$). A reasonable
data collapse occurs for both the toppling size $(s)$ and lifetime
$(t)$ of the avalanche. In the inset, $P(x)s^{\tau_x}$ is plotted
against $s/L^{D_s}$ is plotted in log-normal scale for the same system
sizes.}
\end{figure}

Knowing the values of capacity dimensions, it is now possible to check
the scaling function form given in Eq.\ref{pd} studying the
distribution functions for different system sizes $L$.  The scaling
function form is checked by plotting a scaled distribution
$P(x)L^{D_x\tau_x}$ against the scaled variable $x/L^{D_x}$ for $x\in
\{s,t\}$ in Fig.\ref{fss} following Chessa {\em et
al}\cite{chessa}. For both the properties, toppling size $s$ and
lifetime $t$, a reasonable collapse of data are observed for $L=512,
1024$ and $2048$ in support of the assumed scaling function form in
Eq.\ref{pd}. In the inset of Fig.\ref{fss}, data collapse for toppling
size $s$ is also shown in log-normal scale. The FSS forms assumed here
in RSM for the avalanche properties are then correct. In spite of the
fact that RSM has locally deterministic rule for grain distribution,
it is conservative, its avalanche cluster morphology is almost
compact, it is diffusive, and some of the critical exponents are
similar to that of BTW, it is interesting to note that the scaling
functions in RSM do not follow multi-scaling as in BTW. A sandpile
model with microscopic as well as macroscopic characteristics of a
deterministic model like BTW follows FSS which is characteristics of a
stochastic model like MSM is a new result. This has happened due to
the nature of the rotational constraint which incorporates ``internal
stochasticity'' by changing the state of the critical sites in a time
step during time evolution of the system. Note that, due to the
presence of rotational constraint on the sand flow the toppling
balance of BTW is also broken as already mentioned. Recently, it is
demonstrated by Karmakar {\em et al}\cite{kms} that the scaling
functions obey FSS rather than multi-scaling if toppling imbalance is
introduced in the BTW sandpile model. Existence of FSS in RSM is
possibly due to toppling imbalance as well as ``internal
stochasticity'' in the model.

\subsection{Time Autocorrelation}
\begin{figure}
\centerline{\psfig{file=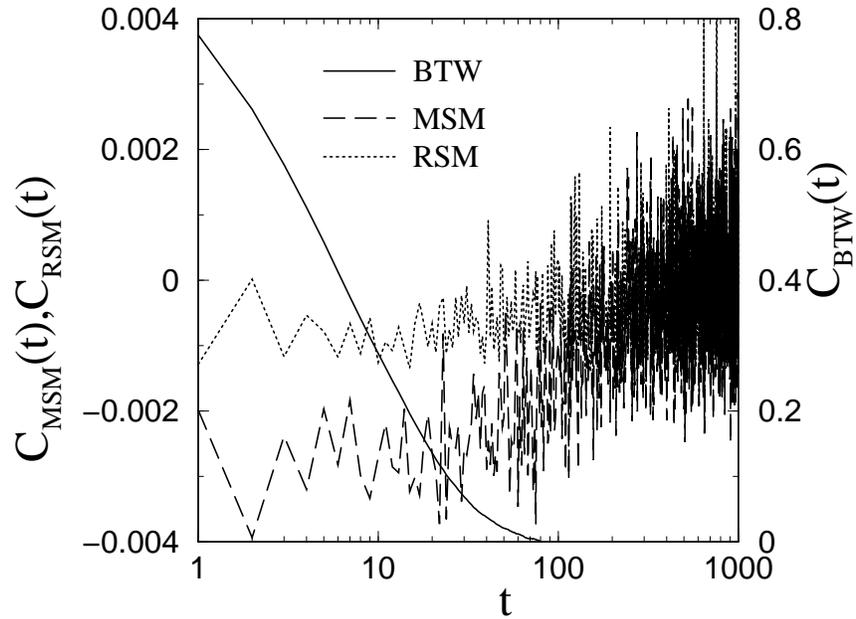, width=0.7\textwidth}}
\medskip
\caption{\label{corr} Time auto correlation function $C(t)$ is plotted
 against time $t$ for BTW (solid line), MSM (dashed line) and RSM
 (dotted line). There is a long range correlation for BTW. The
 toppling waves are uncorrelated for both MSM and RSM. }
\end{figure}

FSS of avalanche properties can be confirmed by studying time
autocorrelation between toppling waves. The time evolution of toppling
dynamics is studied here coarsening the avalanches into a series of
toppling waves\cite{ivash}. Toppling waves are defined as the
number of toppling during the propagation of an avalanche starting
from a critical site $O$ without toppling $O$ more than once. Each
toppling of $O$ creates a new toppling wave. Thus, the total number of
toppling $s$ in an avalanche can be considered as
\begin{equation}
\label{toppl}
s=\sum_{k=1}^m s_k
\end{equation}
where $s_k$ is the number of toppling in the $k$th wave and $m$ is
the number of toppling waves during the avalanche. It is then possible
to generate a wave time series $\{s_k\}$. Following Menech and
Stella\cite{ms}, for a given lattice size $L$, a time autocorrelation
function is defined as
\begin{equation}
\label{corl}
C(t) = \frac{\langle s_{k+t}s_k\rangle -  \langle s_k\rangle^2}{
  \langle s_k^2\rangle - \langle s_k\rangle^2}
\end{equation}
where $t=1,2,\cdots$ and $\langle\cdots\rangle$ represents the time
average. It has already been demonstrated by Menech and
Stella\cite{ms} and Karmakar {\em et al}\cite{kms} that $C(t)$ is long
range for BTW whereas it remains negative initially and then becomes
zero in case of MSM. Thus, the waves in BTW have correlation over a
longer period of time whereas they are uncorrelated in case of
MSM. This is also argued by Menech and Stella\cite{ms} that this
observation is consistent with the fact that the toppling sizs
follow multi-scaling in BTW and FSS in MSM. The time autocorrelation
function $C(t)$ for the toppling waves has also been calculated here
in RSM for a system size $L=1024$ taking time average over $10^6$
toppling waves. $C(t)$ is plotted against $t$ for RSM in
Fig.\ref{corr}. In order to compare the data of RSM with that of BTW,
and MSM, $C(t)$s of these models are also calculated and plotted in
the same figure. It can be seen that there is long range correlation
for BTW and anti-correlation for MSM as expected. The toppling waves
are also uncorrelated here in RSM as in the case of MSM. $C(t)$ is
found negative initially and then becomes zero. It is then consistent
with the fact that the toppling size distribution follow FSS rather
than multi-scaling. The origin of negative autocorrelation in MSM is
the stochasticity. In RMS, the local deterministic toppling rule picks
up certain randomness during the evolution and as a consequence the
toppling wave shows negative auto correlation. Though the rotational
constrain has local correlation, the toppling waves become
uncorrelated because of the ``hidden stochasticity''. It should be
mentioned here that in the cases of MSM and RSM, the sites involved in
a toppling wave may topple more than once unlike in the case of
BTW. Moreover, the toppling numbers as well as the final
configurations of MSM and RSM strongly depend on the sequence of
toppling. As a consequence, Eq.\ref{toppl} may not satisfy always in
RSM and MSM. However, on an average the collection of toppling waves
can be considered as a representation of an avalanche in these models.

\section{Conclusion}
A new two state ``quasi-deterministic'' sandpile model, RSM, is
defined imposing rotational constraint on the sand flow in order to
study the effect of local external bias on self-organized critical
systems. The model has microscopic properties like mass conservation,
open boundary, local deterministicity in sand grain distribution on
toppling as that of BTW. At the same time, the rotational bias
introduces toppling imbalance and certain stochasticity at the
microscopic level as in MSM. The non-equilibrium steady state of RSM
is characterized by power law distribution of avalanche
properties. The avalanche cluster morphology in the steady state
exhibits characteristics of both BTW as well as MSM. RSM is found to
be diffusive in character like other sandpile models. Calculating an
extended set of critical exponents it is found that some of the
exponents are close to that of BTW but different from MSM. The values
of the exponents satisfy the scaling relations among them within error
bars. RSM then belongs to a new universality class. The scaling
function forms are determined. It is found that the scaling functions
obey usual FSS as in the case of MSM rather than multi-scaling as in
the case of BTW. This has been confirmed by negative time
autocorrelation of toppling waves constituting an avalanche. A
sandpile model having certain microscopic as well as macroscopic
features of BTW follows FSS as that of MSM is a new result. The
appearance of FSS in RSM may be due to local toppling imbalance and
``internal stochasticity'' caused by the imposed rotational constraint
on the model.

\vspace{0.75cm}
\noindent
{{\bf Acknowledgment:} The authors thank S. S. Manna, D. Dhar and
  M. A. Munoz for critical comments and several suggestions on the
  manuscript. This work is financially supported by Board of Research
  in Nuclear Sciences, Department of Atomic Energy, India, grant
  no.2005/37/5/BRNS.}

\newpage

\begin{table}
\begin{tabular}{|c|c|c|c|} \hline\hline
  & \multicolumn{3}{|c|}{Models}\\ \cline{2-4}
  Exponent & \multicolumn{1}{|c|}{BTW} &
  \multicolumn{1}{|c|}{MSM}&
  \multicolumn{1}{|c|}{RSM} \\ 
 \hline
  $\tau_s$ & $1.293$   & $1.275$  & $1.224\pm 0.005$  \\
  \hline
  $\tau_a$ & $1.330$  & $1.373$   & $1.334\pm 0.005$  \\
  \hline
  $\tau_t$ & $1.480$  & $1.493$    & $1.389\pm 0.005$  \\
  \hline
  $\tau_l$ & $1.665$  & $1.743$    & $1.667\pm 0.007$   \\
  \hline
  $\gamma_{sa}$ & $1.06$    & $1.23$    & $1.453\pm 0.003$   \\
  \hline
  $\gamma_{at}$ & $1.53$  & $1.35$  & $1.167\pm 0.005$  \\
  \hline
  $\gamma_{al}$ & $2.00$    & $2.00$    & $2.002\pm 0.002$  \\
  \hline
  $\gamma_{tl}$ & $1.32$  & $1.49$  & $1.715\pm 0.005$  \\
  \hline\hline
  
\end{tabular}
\caption{\label{table1} Comparison of critical exponents obtained in
different sandpile models. The values of the critical exponents for
BTW and MSM are taken from Ref.\cite{benhur,usadel}. Some of the
exponents are close to that of BTW and most of them are different from
that of MSM.}

\end{table}
  
\newpage

\end{document}